\documentclass[twocolumn,prb,aps,showpacs,floatfix,superscriptaddress]{revtex4}
\usepackage{graphicx}
\usepackage{verbatim}

\begin{document}

\title{Phonon lineshapes in the vortex state of the phonon-mediated superconductor YNi$_2$B$_2$C}

\author{F. Weber}
\email{frank.weber@kit.edu}
\affiliation{Karlsruher Institut f\"ur Technologie, Institut f\"ur Festk\"orperphysik, P.O.B. 3640, D-76021 Karlsruhe, Germany}
\affiliation{Materials Science Division, Argonne National Laboratory, Argonne, Illinois, 60439, USA}
\author{L. Pintschovius}
\affiliation{Karlsruher Institut f\"ur Technologie, Institut f\"ur Festk\"orperphysik, P.O.B. 3640, D-76021 Karlsruhe, Germany}
\author{K. Hradil}
\altaffiliation[present address: ]{Technische Universit\"at Wien, R\"ontgenzentrum, A-1060 Wien, Austria}
\affiliation{Universit\"at G\"ottingen, Institut f\"ur Physikalische Chemie, Au{\ss}enstelle FRM-II, D-85747 Garching, Germany}
\author{D. Petitgrand}
\affiliation{Laboratoire L$\acute{e}$on Brillouin (CEA-CNRS), CEA-Saclay, F-91911 Gif-sur-Yvette, France}

\begin{abstract}
We present an inelastic neutron scattering study of phonon lineshapes in the vortex state of the type-II superconductor YNi$_2$B$_2$C.  In a previous study [Phys. Rev. Lett. \textbf{101}, 237002
(2008)] it was shown that certain phonons exhibit a clear signature of the superconducting gap $2\Delta$ on entering the superconducting state. Our interest was to find out whether or not the
lineshape of such phonons reflects the inhomogeneous nature of the vortex state induced by a magnetic field smaller than the upper critical field $B_{c2}$ .We found that this is indeed the case
because the observed phonon lineshapes can be well described by a model considering the phonon as a local probe of the spatial variation of the superconducting gap. We found that even at
$B=3\,\rm{T}$, where the inter-vortex distance is less than $300\,$\AA, the  phonon lineshape still shows evidence for a variation of the gap.
\end{abstract}

\pacs{74.25.Kc, 78.70.Nx, 63.20.kd,74.25.Uv}

\maketitle

  \vskip2pc

\section{Introduction}
\label{intro}

Many years ago, Axe and Shirane \cite{Axe73} were the first to show that phonons in phonon-mediated superconductors might exhibit marked changes of their frequency and lineshape if the compound
is cooled below the superconducting transition temperature $T_c$. The effects are particularly pronounced if the phonon frequency is comparable to the energy of the superconducting gap $2\Delta$.
For a long time, it was thought that the phonon lineshape can be well described by a Lorentzian above and below $T_c$ and the superconductivity-induced change of the phonon lineshape is just a
change of the width of the Lorentzian. The first observation of a clearly non-Lorentzian lineshape below $T_c$ was made by Kawano et al. \cite{Kawano96} on the conventional superconductor
YNi$_2$B$_2$C ($T_c=15.2\,\rm{K}$). The low temperature lineshape of certain phonons looked so unusual that the authors of this study thought to have observed the appearance of an additional
excitation below $T_c$. In a later study \cite{Weber08} a variety of strongly non-Lorentzian lineshapes were reported for the same compound. In particular, it was shown that certain phonons
exhibit a step-like increase of their intensity right at the superconducting gap energy $2\Delta$. It was shown that all the non-Lorentzian lineshapes are semi-quantitatively predicted by a
theory proposed by Allen et al. \cite{Allen97}.

In the present study, we address the question what happens to the phonon lineshape if a magnetic field is applied in a type II superconductor. It is quite obvious to expect that the phonon
lineshape will revert to that observed above $T_c$ if the magnetic field is so high as to completely suppress superconductivity, i.e. at fields larger than the upper critical field $B_{c2}$. If,
on the other hand, the applied field is very small, i.e. less than $B_{c1}$, the magnetic field does not penetrate into the superconductor and hence no change is expected for the phonon
properties. Therefore, we deal with fields $B_{c1}<B<B_{c2}$ in the following where the field enters into the superconductor in the form of vortices. In the core of the vortices, the
superconducting gap is completely suppressed but gradually recovers its value without a magnetic field with increasing distance from the vortex core. This means that the vortex state of a type-II
superconductor is inhomogeneous with respect to the size of $2\Delta$. In our experiment, we wanted to find out whether or not the spatial variation of the gap is reflected in the phonon lineshape.

Theory is at present of little help to predict how phonons will respond to the inhomogeneity of the vortex state. In the most basic theory, i.e. in the harmonic approximation, phonons are represented by plane waves traveling through a perfectly periodic atomic lattice without damping. That is to say, not only is the phonon lifetime infinite, but there is also no length scale in the theory. An infinite lifetime is certainly unrealistic due to a variety of reasons, such as phonon-phonon coupling, i.e. anharmonicity, and in metals also electron-phonon coupling. This has led to the development of the quasi-harmonic approximation where phonons do have a finite lifetime, which can be seen in experiment as a broadening of the phonon lines. However, there is as yet no theory to deal with the length scale on which a phonon probes the microscopic properties of the solid. In some cases, one might get an estimate of the length scale by calculating the mean free path $x$ of a phonon from its lifetime and its group velocity via $x=1/\gamma\times v_g$, where $\gamma$ is the inverse phonon lifetime and $v_g=d\omega/d\mathbf{q}$ the group velocity. Both quantities can be obtained from inelastic neutron scattering measurements. For a strong coupling acoustic phonon in Niobium, e.g., one obtains $x = 700\,$\AA \cite{Weber10}, whereas for modes without strong coupling $x$ is of the order of $\mu$m. This approach, however, is not applicable to zone boundary modes and many optic phonons because of their vanishing group velocity, and in particular not to the zone boundary mode studied in our experiment.

The mode chosen in our study was the endpoint of the transverse acoustic (TA) branch along the (110) direction, i.e. at $\mathbf{q} = (0.5,0.5,0)$ (M-point). In a previous study \cite{Weber08} we had found that the lineshape of this phonon shows a very steep intensity increase right at the energy of the superconducting gap $2\Delta$ and therefore, a smearing or a shift of the gap could be easily detected in measurements in an applied field. Magnetic field dependent measurements have already been reported in Ref. \onlinecite{Kawano96}, using a TA phonon at $\mathbf{q} = (0.55,0,0)$ (see Sect.~\ref{discussion}). The lineshape of this phonon does show pronounced changes on entering the superconducting state and therefore lends itself to a study of the influence of a magnetic field as well. However, the superconducting energy gap cannot as easily be extracted from the lineshape of this phonon as in the case of the M-point phonon \cite{Weber08}, and all the more any shift or smearing
of the gap.

In our experiment, we measured the evolution of the phonon lineshape of the M-point phonon as a function of applied magnetic fields $0\,\rm{T} \le B \le 9\,\rm{T}$ ($B_{c2} = 9.3\,\rm{T}$ at $T =
2\,\rm{K}$ [\onlinecite{Shulga98}]). We show that at small fields, i.e. large inter-vortex distances, the phonon lineshape clearly reflects the spatial variation of the superconducting gap in the
crystal lattice. With increasing field strength, the superconductivity-induced deformation of the phonon lineshape is more and more reduced by the high density of vortices, which makes it
progressively difficult to obtain an unambiguous result. Our data obtained at $B=3\,\rm{T}$, i.e. inter-vortex distances less than $300\,$\AA$\,$, still show evidence for a spatial variation of
the gap, but in a less convincing way than at low fields.

\section{Experimental}
 \label{experimental}

  \begin{figure}
   \includegraphics[width=0.9\linewidth]{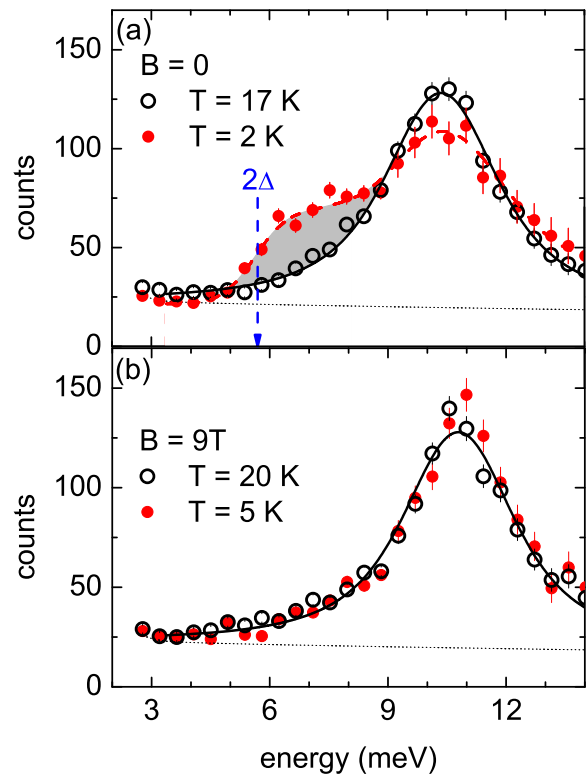}
   \caption{(color online) Raw data obtained on the TAS 4F2 of a transverse acoustic phonon at $\mathbf{Q} = (0.5, 0.5, 7)$ (run 1). Energy scans were measured at \textit{(a)} $B = 0$ and \textit{(b)} $9\,\rm{T}$ well below and above the superconducting transition temperature in zero field $T_c = 15.2\,\rm{K}$. The vertical dashed line in \textit{(a)} denotes the value of the superconducting gap at $T = 2\,\rm{K}$. Solid lines are fits to the data using a Lorentzian convoluted with the experimental resolution on an estimated background (dotted lines). The dashed line in \textit{(a)} is a guide to the eye. The size of the shaded area in \textit{(a)} was used to measure the strength of the lineshape deformation (see text and Fig.~\ref{fig_4})}
   \label{fig_1}
  \end{figure}

The neutron scattering experiments were performed on the cold triple-axis spectrometer 4F2 at the ORPHEE reactor at LLB, Saclay, (two runs, 1 and 2) and on the thermal triple-axis spectrometer PUMA at the research reactor FRM II in Munich. Double focusing pyrolytic graphite monochromators and analyzers were employed in both cases. At PUMA a fixed analyzer energy of $14.7\,\rm{meV}$ allowed us to use a graphite filter in the scattered beam to suppress higher orders. Due to technical limitations of the double focusing monochromator on 4F2, scans could be done using a fixed analyzer energy of $14.7\,\rm{meV}$ only up to an energy transfer of $9\,\rm{meV}$, whereas for larger energy transfers the incident energy was kept constant at $E_i = 23.7\,\rm{meV}$. Therefore, the data had to be corrected for the $\mathbf{k}_f$ dependence of the analyzer efficiency. On PUMA, data were collected for fields $0.25\,\rm{T} \le B \le 1.0\,\rm{T}$, whereas data for higher fields were collected on the 4F2 spectrometer. The magnetic field was applied perpendicular to the $(hh0)-(00l)$ scattering plane, i.e. along the $(1\overline{1}0)$ direction. Changes in the applied magnetic field were always done in the non-superconducting state, i.e. at $T>T_c = 15.2\,\rm{K}$ (field cooled procedure). For each round of experiments zero field data were taken as well at base temperature and at a temperature slightly above the superconducting transition temperature $T_c = 15.2\,\rm{K}$. In order to achieve an adequate counting statistics, individual scans were repeated several times. Since it was clear that a spatial variation of the superconducting gap would show up at energies below $\approx 8\,\rm{meV}$, a large fraction of the total counting time for each scan was allotted to the low energy region. Unavoidably, this strategy led to relatively large statistical errors at the upper end of the energy range investigated.

The single crystal sample of YNi$_2$B$_2$C weighing $2.26\,\rm{g}$ was grown by the floating zone method using the $^{11}$B isotope to avoid strong neutron absorption \cite{Souptel}. The specimen was mounted in a $10\,\rm{T}$ Oxford cryostat at 4F2 and a cryogen-free $7\,\rm{T}$ magnet on PUMA employing a standard closed-cycle refrigerator allowing measurement down to $T = 2\,\rm{K}$ and $3\,\rm{K}$, respectively. Measurements were carried out in the $(hh0)-(00l)$ scattering plane. The wave vectors are given in reciprocal lattice units ($rlu$) of ($2\pi/a$ $2\pi/b$ $2\pi/c$), where $a = b = 3.51\,$\AA$\,$ and $c = 10.53\,$\AA.

\section{Results}
\label{results}
  \begin{figure}
   \includegraphics[width=0.9\linewidth]{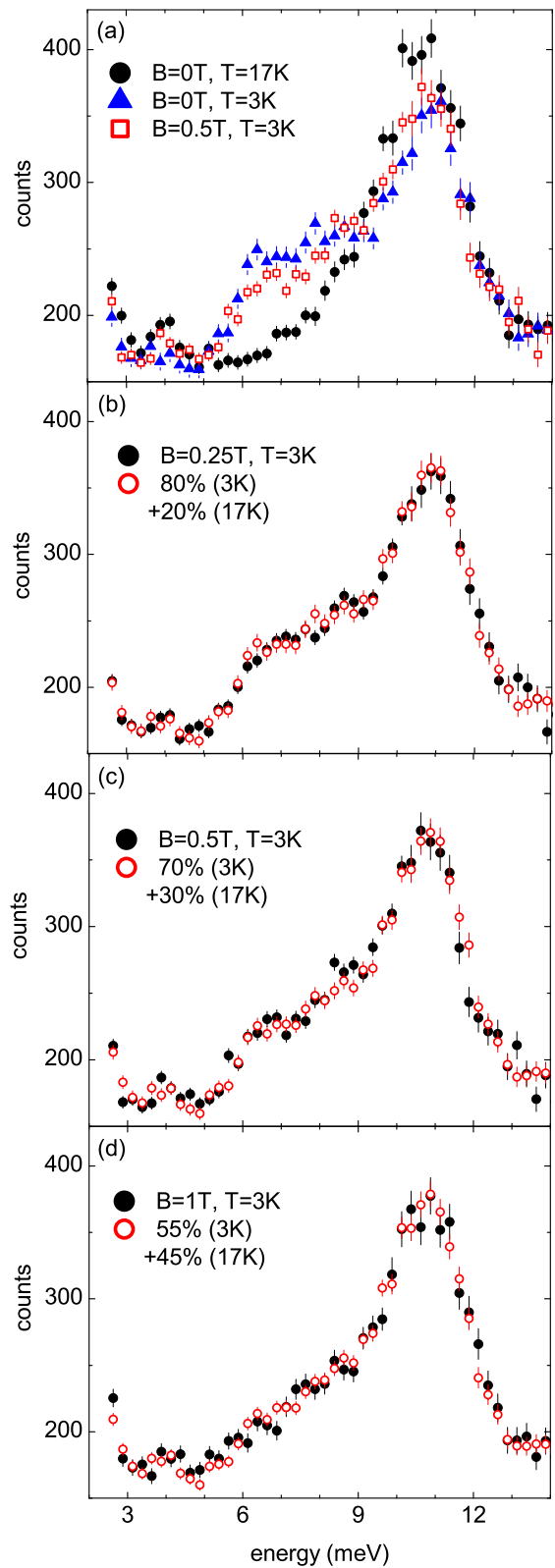}
   \caption{(color online) Raw data of energy scans of the transverse acoustic phonon at $\mathbf{Q} = (0.5, 0.5, 7)$ obtained on the TAS PUMA. \textit{(a)} Comparison of zero field and $B = 0.5\,\rm{T}$ data. Energy scans at \textit{(b)} $B = 0.25\,\rm{T}$, \textit{(c)} $0.5\,\rm{T}$ and \textit{(d)} $1.0\,\rm{T}$ at $T = 3\,\rm{K}$ (dots). The open circles in \textit{(b)-(d)} correspond to superpositions of the raw data at $T = 3\,\rm{K}$ and $17\,\rm{K}$ in zero field (see panel \textit{(a)}) with the ratios given in the legend of each panel.}
   \label{fig_2}
  \end{figure}

  \begin{figure}
   \includegraphics[width=0.9\linewidth]{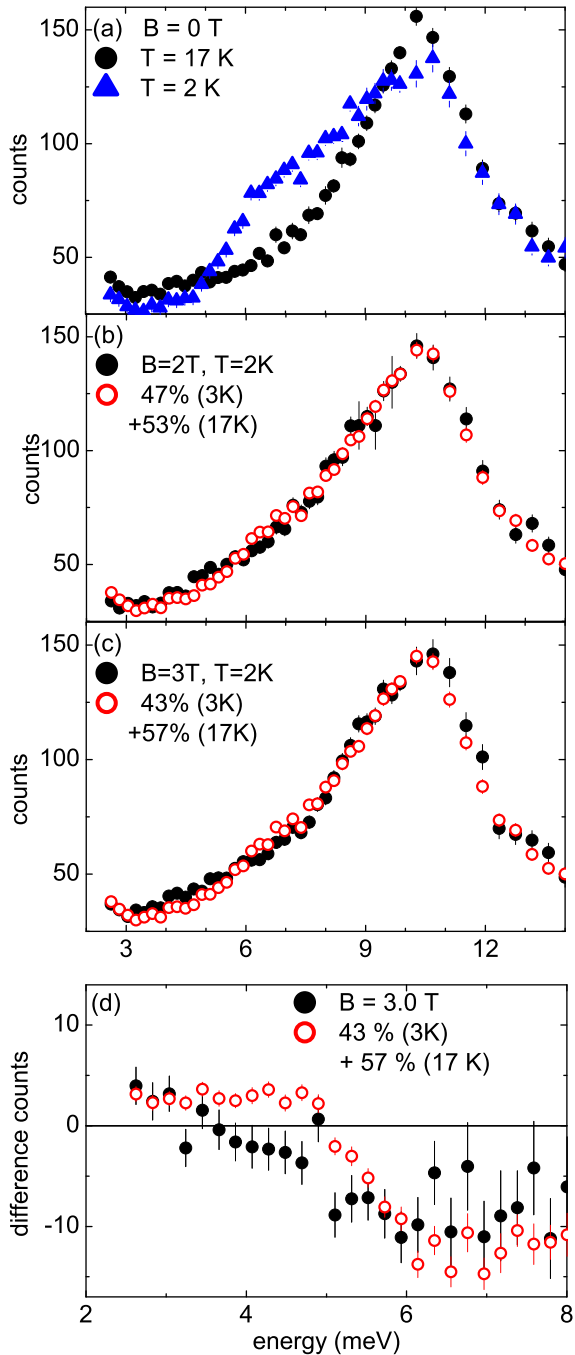}
   \caption{(color online) Raw data of energy scans obtained on the TAS 4F2 of a transverse acoustic phonon at $\mathbf{Q} = (0.5, 0.5, 7)$ (run 2). Energy scans were measured in \textit{(a)} zero field, \textit{(b)} $B = 2.0\,\rm{T}$, \textit{(c)} $3.0\,\rm{T}$ at $T = 2\,\rm{K}$ and $T = 17\,\rm{K}$ (only zero field). \textit{(d)} Enlargement of the low energy region of panel \textit{(c)} including data taken at $T = 12\,\rm{K}$ in zero field (open squares). The open circles in \textit{(b)-(c)} correspond to superpositions of the raw data at $T = 3\,\rm{K}$ and $17\,\rm{K}$ in zero field with the ratios given in the legend for each panel. \textit{(d)} Difference counts between normal and superconducting phase at small energy transfers, for which the data shown in \textit{(c)} (same labeling) were subtracted from zero field data at $T = 17\,\rm{K}$}
   \label{fig_3}
  \end{figure}

  \begin{figure}
   \includegraphics[width=0.95\linewidth]{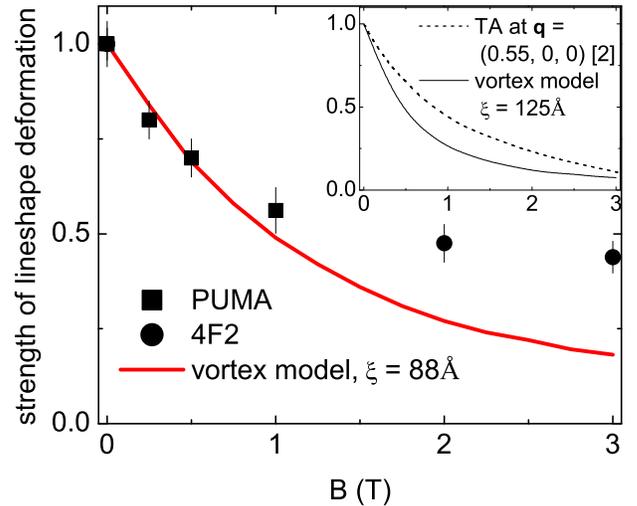}
   \caption{(color online) Strength of the phonon lineshape deformation normalized to its maximal value at low temperature as function of magnetic field. The solid line is based on the vortex model as described in the text. Inset: The dashed line represents the magnetic field dependent lineshape deformation of the acoustic phonon at $\mathbf{q}=(0.55,0,0)$ (reproduced from Fig. 4 of Ref.~\onlinecite{Kawano96}, see text). The solid line was calculated from our vortex model, but with a larger coherence length of $\xi=125\,$\AA$\,=\sqrt{2}\times 88\,$\AA$\,$ corresponding to the reduction of the upper critical field by a factor of $2$ from  from $B_{c2}=9.3\,\rm{T}$ at $T=2\,\rm{K}$ to $B_{c2}=4.7\,\rm{T}$ at $T=5.5\,\rm{K}$.}
   \label{fig_4}
  \end{figure}

  \begin{figure}
   \includegraphics[width=0.92\linewidth]{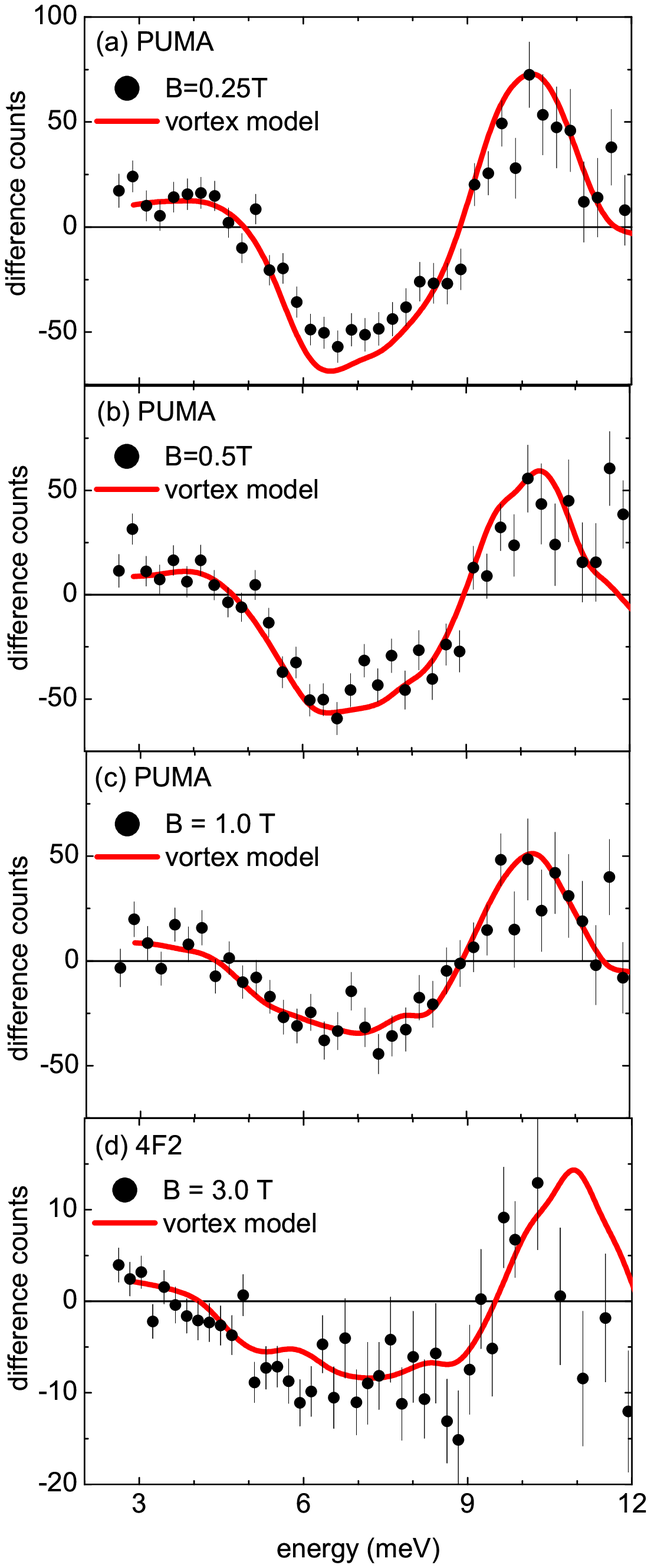}
   \caption{(color online) Difference between neutron counts in the normal and superconducting state (dots) at B = 0.25 \textit{(a)}, 0.5 \textit{(b)}, 1.0 \textit{(c)} and 3.0 T \textit{(d)} compared to predictions based on a vortex model (solid lines, see text). Experimental data were taken on the TAS's PUMA \textit{(a)(b)(c)} and 4F2 \textit{(d)}.}
   \label{fig_5}
  \end{figure}

Energy scans performed on the TAS 4F2 in zero magnetic field and $B = 9\,\rm{T}$ are shown in Fig.~\ref{fig_1}. The zero field data taken at $T=2\,\rm{K}$ exhibit the abrupt increase of the
phonon intensity at the superconducting gap energy $2\Delta = 5.7\,\rm{meV}$ as expected from our previous measurements \cite{Weber08} albeit this feature is slightly washed out when compared to
the results reported in Ref.~\onlinecite{Weber08} due to a slightly lower energy resolution of this instrument. On the other hand, the data taken on the TAS PUMA show again a very sharp feature at
$E=2\Delta$, which allowed us to unambiguously detect the influence of magnetic fields as small as $0.25\,\rm{T}$ (Fig.~\ref{fig_2}).

We do not find any statistically significant difference between scans at $T = 20\,\rm{K}$ and $5\,\rm{K}$ at $B = 9\,\rm{T}$ (Fig.~\ref{fig_1}b). We note that the applied field of $9\,\rm{T}$ exceeds the upper critical at $T = 5\,\rm{K}$ \cite{Shulga98}. This result does not come as a surprise but just demonstrates that the distorted phonon lineshape is linked to the presence of the superconducting phase and the corresponding energy gap. We now turn to the results depicted in Fig.~\ref{fig_2}, i.e. obtained at fields $B \le 1\,\rm{T}$. The scans performed at $B=0.25\,\rm{T}$ or $0.5\,\rm{T}$ show the same step-like increase of the intensity at $E=5.7\,\rm{meV}$ as that performed in zero field albeit the step height is significantly reduced. The lineshape in these scans can be very well reproduced by a superposition of the lineshapes observed in zero field at $T=3\,\rm{K}$ and $17\,\rm{K}$. In other words, the lineshape can be very well described by the assumption that the phonons sample volume fractions with zero or the full gap, respectively. The same model is fairly successful for the $1\,\rm{T}$-data as well, although a close look reveals that the step at $E=5.7\,\rm{meV}$ is somewhat smeared out. A more elaborate model to describe these data as well as those obtained at higher fields will be presented in the next section. At higher fields (Fig.~\ref{fig_3}), a superposition of zero field data taken at $T = 2\,\rm{K}$ and $17\,\rm{K}$, respectively, gives only an unsatisfactory account of the observed lineshape. This can be seen more easily comparing the superconductivity-induced changes, i.e. the difference between neutron counts in the normal and superconducting state, where each of the two data sets shown in Fig.~\ref{fig_3}c were subtracted from the counts measured at $T = 17\,\rm{K}$  (Fig.~\ref{fig_3}d). Such a discrepancy had to be expected from the fact that in high fields, the gap will nowhere reach the full zero field value.

In order to establish a quantitative relationship between the strength of the superconductivity-induced lineshape deformation and the magnetic field, we evaluated the integrated difference of neutron counts between the lineshapes above $T_c$ and at very low $T$ as shown in Fig.~\ref{fig_1}a as gray shaded area. The results of this analysis are depicted in Fig.~\ref{fig_4}. Somewhat surprisingly, the reduction of the lineshape deformation depends in a highly non-linear way on the magnetic field. Possible reasons for this observation will be discussed in the next Section.

\section{Discussion}
\label{discussion}

We now want to discuss our results in the light of what is known
 for the vortex state of a type-II superconductor such as YNi$_2$B$_2$C. Here, an applied magnetic field B enters the superconducting sample in the form of flux lines, so-called vortices, for $T < T_c$ and $B_{c1} < B < B_{c2}$. The superconducting gap in the center of a vortex is zero and the coherence length $\xi$ defines the radius around a vortex core where $\Delta$ reaches half of its maximum value. If a phonon would be a completely local probe, the experimental spectral weight distribution in the vortex phase should reflect the volume fractions having specific gap values. Starting from this assumption, we calculated the volume fractions with specific gap values using a value of $\xi=88\,$\AA$\,$ as reported in the literature \cite{Eskildson97} and the functional form of the dependence of the gap on the distance from the vortex core as published in textbooks \cite{Tinkham}. The magnetic field enters into this calculation via the inter-vortex distance $a_{vortex}$, which is given by \cite{Tinkham,Sakata00}
\begin{equation}\label{equ1}
a_{vortex}=\frac{5\times 10^{-8}\,\rm{m}}{\sqrt{B/\rm{T}}}
\end{equation}
In the next step, we summarized the contributions from each volume fraction using the lineshapes observed for specific gap values in zero field in a previous experiment \cite{Weber08}. We note that these data, although collected with a different set-up, can be very well transferred to the actual experiment after proper normalization (see Appendix).

Results obtained in this way are shown in Fig.~\ref{fig_5}. For the sake of an easier assessment of our approach, we did not plot the observed and calculated intensity distributions but instead the superconductivity-induced lineshape deformations for various field values. Apparently, our approach works quite well in the whole field range explored in our experiment. The good agreement even for $B=3\,\rm{T}$ is somewhat surprising in view of the fact that our model is oversimplified for short inter-vortex distances: it is known from the literature that the distance dependence of the superconducting gap from the vortex core derived for isolated vortices is significantly modified for high vortex densities \cite{Brandt03}. More specifically, it is expected that the gap size is more strongly reduced half-way to the next vortex core than calculated from the functional form for isolated vortices. Attempts to include such an effect into our model led to a serious disagreement with experiment. Later, we became aware that the good agreement between model and experiment at $B=3\,\rm{T}$ (and $2\,\rm{T}$ as well) is somewhat accidental because our model is somewhat unrealistic in another aspect: it assumes a field-independent value of the vortex diameter $\xi$ whereas investigations by muon spin resonance have provided evidence for a drastic reduction of $\xi$ at high magnetic fields \cite{Ohishi00}. However, taking the value of $\xi = 40\,$\AA$\,$ reported in \cite{Ohishi00} for $B=3\,\rm{T}$ at face value leads to a massive overestimation of the superconductivity-induced lineshape deformation: the effect should be as strong as calculated for $\xi =88\,$\AA$\,$ at $B=0.5\,\rm{T}$ (Fig.~\ref{fig_5}b), which is evidently incorrect. This discrepancy might be due to the fact the magnetic fields were applied along $(001)$ in Ref.~\onlinecite{Ohishi00} but along  $(1\overline{1}0)$ in our work. We did not try to deduce a field dependence of $\xi$ from our data in order not to overstretch the information contained in our data, but concluded that there is insufficient knowledge of the properties of the vortex state in YNi$_2$B$_2$C at high fields for a realistic simulation of the phonon lineshape.  An important qualitative conclusion can be drawn from our high field data nevertheless: the pronounced smearing of the feature related to the superconducting energy gap observed at $B=3\,\rm{T}$ indicates that the phonon lineshape reflects the inhomogeneous nature of the vortex state even for an inter-vortex distance as small as $300\,$\AA.

A variation of the vortex diameter as function of the applied field from $\xi =100\,$\AA$\,$ for very small fields to $40\,$\AA$\,$ at high fields seems to be unrealistic but on the other hand, our results do confirm such an effect qualitatively as evident from the highly non-linear influence of the magnetic field on the lineshape deformation. The magnetic field dependence of the superconductivity-induced effect strength based on the structure of an isolated vortex is depicted as solid line in Fig.~\ref{fig_4}. Taking into account that in regions with overlapping vortices the maximum gap value is further reduced \cite{Brandt03} the effect should fall off with increasing magnetic field even faster. However, our results show a larger effect for $B \ge 1\,\rm{T}$. This discrepancy can be qualitatively understood by a reduction of $\xi$ causing generally increased gap values compared to our model and a corresponding stronger deformation of the phonon lineshape.

As mentioned in the introduction, the magnetic field dependence of the lineshape of another strong coupling phonon in YNi$_2$B$_2$C was already investigated in Ref.~\onlinecite{Kawano96}. We pointed out that a shift or a smearing of the superconducting gap cannot be easily derived from the lineshape of this phonon, because an evaluation of the superconducting gap relies heavily on the theory proposed by Allen et al.\cite{Allen97}, which, however, describes the observed effects only semi-quantitatively. A more summary information can be obtained nevertheless: in the inset of Fig.~\ref{fig_4} we compare the reported magnetic field dependent intensity of what the authors of Ref.~\onlinecite{Kawano96} considered as a new peak with the strength of the lineshape deformation computed from our model outlined above. In order to account for the  reduced value of the upper critical field at $T=5.5\,\rm{K}$, the temperature at which the experiments reported in Ref.~\onlinecite{Kawano96} were performed, we used a value of  $\xi=125\,$\AA$\,$ instead of $88\,$\AA\footnote{Although the temperature dependence of $\xi(T)$ was considered theoretically\cite{Buzea01}, there are no experimental measurements of $\xi(T=5.5\,\rm{K})$ for YNi$_2$B$_2$C.}. Again, we find a weaker reduction of the effect as function of magnetic field than expected for a constant value of $\xi$. Thus, the results for the TA phonon at $\mathbf{q}=(0.55,0,0)$ corroborate our conclusion that there it is indeed a decrease of $\xi$ as a function of magnetic field.


\begin{figure}
   \includegraphics[width=0.9\linewidth]{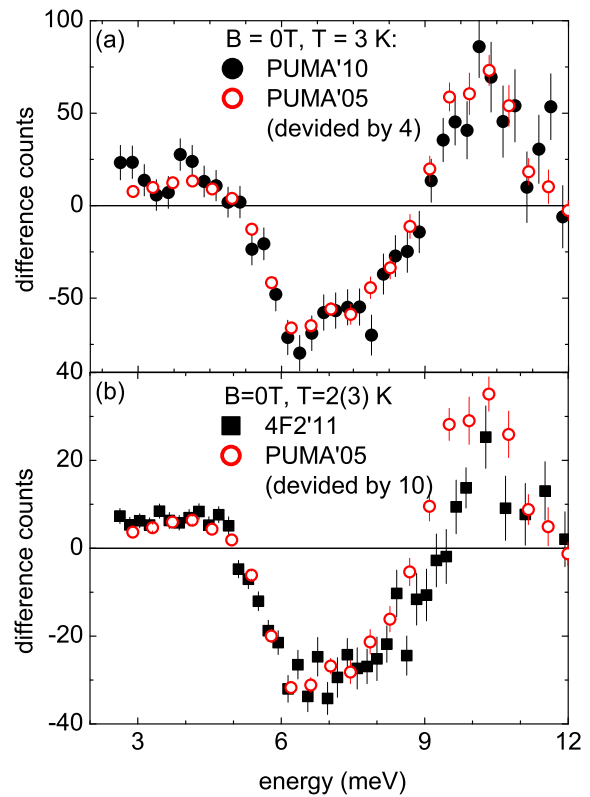}
   \caption{(color online) Difference between neutron counts in the normal and the superconducting state in zero magnetic field as observed in three different experiments. All data were properly normalized. \textit{(a)} Data were taken on the PUMA spectrometer in the current investigation (filled symbols, with the sample mounted inside a superconducting magnet) and a previous experiment (open symbols, no magnet). \textit{(b)} Data were taken in the early experiment on PUMA (open symbols) or the 4F2 spectrometer with the sample inside a 10T-magnet (full symbols).}
   \label{fig_6}
  \end{figure}

\section{Conclusions}

We have reported the magnetic field dependence of the lineshape of a certain phonon in YNi$_2$B$_2$C, which exhibits a clear signature of the superconducting gap when the sample is cooled in zero field below $T_c$. The observed lineshapes can be very well reproduced by a model based on the assumption that the phonon acts as a local probe of the inhomogeneous gap distribution in the vortex phase. The model is quite convincing for fields up to $B \le 1\,\rm{T}$. Even at $B = 3\,\rm{T}$, where the inter-vortex distance is less than $300\,$\AA, the strong smearing of the superconductivity-induced feature indicates that the phonon resolves the gap values locally. A quantitative comparison between calculation and experiment at high fields is complicated by the fact that the vortex diameter seems to be field dependent in YNi$_2$B$_2$C. Our analysis corroborates conclusions drawn from muon spin resonance data\cite{Ohishi00} that the vortex diameter shrinks considerably in high magnetic fields in YNi$_2$B$_2$C.

\begin{acknowledgments}
We acknowledge valuable discussions with R. Heid. Work at Argonne was supported by U.S. Department of Energy, Office of Science, Office of Basic Energy Sciences, under contract No. DE-AC02-06CH11357.
\end{acknowledgments}
\vspace{0.8cm}
\section{appendix}

As explained in our discussion, the curves shown in Fig.~\ref{fig_5} were calculated using observed phonon lineshapes for specific gap values. More precisely, we used zero field data taken at different temperatures whereby the sharp step in the intensity distribution allowed us to associate each temperature $T \le T_c$ with a distinct gap value. However, the limited beam time assigned to the experiments in a magnetic field did not allow us to collect these data with the set-up used for the results reported in this paper. Therefore, we resorted to use data from a previous experiment \cite{Weber08} performed without a magnet. We note that this experiment involved very long counting times and therefore led to good counting statistics. In Fig.~\ref{fig_6} we demonstrate that the superconductivity-induced deformation of the phonon lineshape observed in the experiments on the PUMA spectrometer is very much the same in different sample environments
after proper normalization. The agreement between the data taken in 2005 on the PUMA spectrometer and those in 2011 on the 4F2 spectrometer is not as good but still very satisfactory up to an energy transfer of about $8\,\rm{meV}$. At larger energy transfers, there are systematic deviations between the PUMA and the 4F2 data, probably because the 4F2 instrument did not allow us to use a constant $\mathbf{k}_f$ throughout the upper part of the scan. Therefore, our model calculations for the results obtained on 4F2 spectrometer sit on firm ground only up to about $8\,\rm{meV}$.

\end{document}